\documentclass[pre,amsmath,showpacs,superscriptaddress,floatfix]{revtex4}

\usepackage{multirow}
\usepackage{rotating}
\usepackage{color}

\begin{document}

\title{Teach Network Science to Teenagers}

\author{Heather A. Harrington} \affiliation{Division of Molecular
  Biosciences, Imperial College London,
  London, SW7 2AZ, UK}

\author{Mariano \surname{Beguerisse D\'iaz}} \affiliation{Department
  of Mathematics, Imperial College London, London,
  SW7 2AZ, UK}

\author{M. Puck Rombach} \affiliation{Oxford Centre for Industrial and
  Applied Mathematics, Mathematical Institute, University of Oxford,
  OX1 3LB, UK}
  
\author{Laura M. Keating} \affiliation{Oxford Centre for Industrial and
  Applied Mathematics, Mathematical Institute, University of Oxford,
  OX1 3LB, UK} 

\author{Mason A. Porter} \affiliation{Oxford Centre for Industrial and
  Applied Mathematics, Mathematical Institute, University of Oxford,
  OX1 3LB, UK} \affiliation{CABDyN Complexity Centre, University of
  Oxford, Oxford, OX1 1HP, UK}







\begin{abstract}

We discuss our outreach efforts to introduce school students to network science and explain why networks researchers should be involved in such outreach activities.  We provide overviews of modules that we have designed for these efforts, comment on our successes and failures, and illustrate the potentially enormous impact of such outreach efforts.

\end{abstract}

\maketitle


\section{Introduction and Motivation}\label{intro}

The World Wide Web, friendship networks at schools, and transportation networks are all complex interconnected systems that are based not only on components but also on interactions between components \cite{newman2010}. To try to understand such systems, one can use network science (i.e., the science of connectivity).  The notion of networks has become a mainstream part of everyday life, and scholars from sociology, physics, computer science, mathematics, and many other disciplines have developed theoretical tools and performed empirical analyses to obtain profound insights on a diverse array of natural and designed phenomena.  

Networks are fundamentally interesting to children and mathematicians alike. They are accessible on an intuitive level using visualisations and simple calculations, and we believe that scholars and the lay public are interested in networks for essentially the same reasons. The ubiquity of networks offers a glimpse into the type of questions that scholars consider when studying networks, and many of these questions---and, at an intuitive level, even their solutions---can already be appreciated by children with just a primary school education.  Most children have regular contact with networks (through Facebook, Google, etc.), and this makes it possible to impart intuitive understanding of key concepts from network science in short outreach sessions. Importantly, because networks are omnipresent, they also provide an ideal means (a gateway drug) to motivate students to pursue mathematics and science in greater depth. 

There are many ways to help inspire children to study mathematics and other scientific disciplines.  For example, science books written for adolescents can be extremely valuable \cite{mason:quantum:2013}.  Other resources, such as the ``Being a Professional Mathematician" project, demonstrate that science is fundamentally a human endeavor \cite{mann-being}.  Still others, such as the Ig Nobel Prizes, illustrate the sheer joy of scientific discovery \cite{air-website}.  In this editorial, we will advocate outreach activities, which provide a medium for students to interact with specialists and learn first-hand what scientists do \cite{saab:hypothesis:2010,dwyer:ams:2001,bass:ams:2001}.  

We have developed a program of outreach activities to teach school students about network science.  We also hope to pass along some of our enthusiasm for networks and for mathematics (and science) more generally. We concentrated on ages 13--16, as the younger members of this age group in the United Kingdom have not yet chosen their subject specializations.  We conducted our outreach activities both at local schools and at Somerville College in Oxford, and they lasted anywhere from a couple of hours to half a day.  At each outreach event, we presented various aspects of network science and discussed more general issues such as studying mathematics at universities.  

An outreach setting makes it possible to discuss with students what scientists actually do. Rather than showing a result after it has been packaged and tied up nicely with a bow, we guide the students towards their own discovery of network principles.  This contrasts with the usual (and perhaps somewhat dry) material that the students are used to seeing, and we hope that it inspires them to further study.  We also try to illustrate that mathematics---not just networks---is a fundamental part of the students' lives by virtue of its connection to known everyday experiences like using Google or traversing a network of roads.  

Our outreach activities highlight a side of mathematics that is crucial but likely unfamiliar to the students: curiosity to understand a vexing (and perhaps ambiguous) question is the main drive, and the blood, sweat, and tears of repetitive calculation (which come later) can temporarily be pushed aside in favor of a big picture. The distinction between repetition and creative thinking about science is common at universities, but Richard Feynman's ``pleasure of finding things out" \cite{feynman-pleasure} can---and should---be available for people of all ages.

The rest of this paper is organized as follows.  We present our outreach activities in Section \ref{sec2}.  We introduce our modules in Section \ref{sec3} (and discuss them in detail in appendices). We examine the successes and challenges of our outreach activities in Section \ref{sec5} and provide arguments in favor of teaching network science to teenagers in Section \ref{sec6}. In Section \ref{sec7}, we present an outlook and encourage others to participate in and develop similar activities.  In Supplementary Online Material (SOM), we include lesson materials for five of the modules, and we encourage you to use and adapt them for your own activities.


\section{Our Outreach Activities}\label{sec2}

We ran events in which students visited us at Somerville College in Oxford and others in which we visited them. The number of students varied significantly (from about 5 to about 50) from one event to another, and their ages ranged from 13 to 16.  After introducing ourselves to the students and teachers, we gave a short introductory talk \footnote{For events at Somerville College, we also included a short introduction to the University of Oxford, studying mathematics and science at a university, and Somerville College before we started discussing network science.} in which we defined ``network" and some other relevant terms. We also introduced ``network science" as the science of connectivity, showed several diverse examples of networks, and gave tantalizing hints as to how investigating network structure can provide information about dynamics or function.  We then split the students into smaller groups (e.g., 50 students into 3 groups) in breakout sessions so that they could delve deeper into a specific topic.  

An example of an introductory talk on networks is available at \cite{mason-intro-talk}.  This introductory talk takes 20--30 minutes.  (We varied the amount of time that was spent discussing the various examples.) The presentation gives the definition of a network and shows how to represent a network as an adjacency matrix \footnote{Some of the students had seen matrices before, but others had not.} using a small example.  The speaker indicates different types of networks (e.g., unweighted, weighted, directed, etc.) and asks the students to think about how the matrix representation can be generalized for the different cases.  The presentation then includes numerous examples, which are introduced via pretty pictures 
(just like in other talks for general audiences), and comments on how it relates to the students' experiences.  Examples that we discussed include London's metropolitan transportation network (``The Tube"), Facebook friendships, food webs, networks in online role-playing games, and Web pages connected by hyperlinks. The talk also purposely introduces a \emph{small} amount of jargon---such as ``node", ``edge", ``small world", and ``degree"---to provide some terminology to facilitate discussions in the breakout sessions.  The introductory talk also includes hints of the mathematics under the hood, but it focuses predominantly on using broad brush strokes to introduce a few important ideas.

After the introductory talk, we break out into sessions for students to explore in detail in smaller groups.  (See Section \ref{sec3} for the topics and the appendices and SOM for lots of detail.)  Each of these sessions, which had 3--5 students per volunteer, was led by 1--2 people with 1--2 others helping.  In many cases, the school teachers participated actively and were extremely helpful.  
Depending on the event, our sessions lasted 30--45 minutes, and each student participated in 1--2 such sessions \footnote{As we became more experienced with our efforts, we concluded that shorter sessions were a better format than longer sessions for most of the students.}.  On many occasions, we also asked the students to present their findings to the other groups.  Naturally, our sessions had broader aims beyond discussions of specific ideas from network science: we wanted the students to think independently (and in small groups) and to investigate difficult, open-ended problems instead of problems with neat answers (to which they were more accustomed). We wanted to give the students a sense of how a scientist might tackle a problem or at least to give them something closer to what a university student might experience.


We had two types of outreach events: (1) ones in which students and teachers visited us at Somerville College in Oxford, and (2) ones in which we visited the schools.  For the events at Oxford, the logistics (including essential items like food, travel, and coffee) were arranged and run by Amy Crosweller, Somerville's Access and Communications Officer.
Amy also helped recruit Somerville undergraduates (usually people studying mathematics) to assist us by giving tours of the College, answering questions about what it is like to be an undergraduate at University of Oxford and in Somerville College, and occasionally even acting as helpers for the breakout sessions.  In one of the events in Oxford, we had a panel discussion about careers that included a volunteer from Google.  

Hosting the students in Somerville College had several advantages---it made it possible to draw students from different schools to the same event, it allowed more control and knowledge of the local facilities (flip charts, markers, etc.), and it gave us access to friendly and helpful undergraduates.  However, we ultimately decided that the ``travelling road show" format was more effective.  This format allows us to work with students who are farther away geographically, as only nearby schools came to the Oxford events in practice even though we had money to offer them accommodation.  Traveling to the schools is also important for attempting to work with students from schools who do not typically send students to University of Oxford (or, in some cases, even consider that as a possibility that's actually on the radar).  It also makes it easier to have an event on a weekday rather than a weekend \footnote{On one occasion, we tried having a weekday event in Oxford, but attendance was sparse.}, as the students can attend this special event rather than a normal class for an hour or two.

We varied details of the event components to accommodate school schedules, road-show versus Oxford events, number of volunteers, number and type (e.g., age and quality) of students who would be working with us, and lessons we learned regarding what seemed to work and what didn't.  For example, in one event held in Somerville College, we worked with the same set of students for 6 hours \footnote{In retrospect, this was probably too long for them.}: we held two 45-minute sessions (with parallel breakout modules during each session), discussions over lunch, and more. Our other events were shorter.  In our visits to schools, we experienced a wide range of abilities and behavior among the students. 
Sometimes we had students from the same grade level and other times there was a mixture. The latter was particularly enjoyable, as it appears to be unusual for students of different ages to work with each other on equal footing.  Schools varied on whether they asked us to work with their top students, standard students, or underachievers.

Our outreach events benefited tremendously from numerous volunteers---including professors, postdoctoral scholars from University of Oxford and Imperial College London, doctoral students from several different disciplines (predominantly mathematical scientists but also several biologists), visiting researchers, undergraduates, staff members, and the students' own teachers.  
The Web pages
\url{http://www.some.ox.ac.uk/191-5717/all/1/Somerville_tutor_helps_students_harness_the_science_behind_social_networks.aspx}
and
\url{http://blogs.some.ox.ac.uk/access/2012/04/30/motivating-maths-pupils-and-reflecting-on-the-latest-sutton-trust-survey/}
contain descriptions of our outreach efforts.





\section{Lesson Materials} \label{sec3}


Our session plans are designed for modules that last about 30-45 minutes, but they can be adapted readily for other formats. For example, we have occasionally merged multiple modules into a single module that covers a broader set of topics.  The modules are interactive, and each one allows participating students to learn about one or more areas (or applications) of network science.  

In the appendices, we describe the following modules:
\begin{enumerate}
\item{Appendix \ref{lesson1}: climate change and food webs;}
\item{Appendix \ref{lesson2}: small worlds and social networks;}
\item{Appendix \ref{lesson3}: disease spread and vaccination strategies;}
\item{Appendix \ref{lesson4}: Google's PageRank algorithm;}
\item{Appendix \ref{lesson5}: coloring maps and other puzzles as an introduction to proving theorems in graph theory;}
\item{Appendix \ref{lesson6}: why your friends have more friends than you do;}
\item{Appendix \ref{lesson7}: structural balance in networks.
 }
\end{enumerate}
Modules 1--5 were our most successful ones, though we're sure you'll have ideas for how to improve them.  We discuss all seven of the above modules in the appendices and go into greater depth for Modules 1--5 in the SOM.








\section{Successes and Challenges} \label{sec5}

Our outreach events have been largely successful, and (unsurprisingly)
more iterations have led to greater polish and greater success.  In
this section, we'll highlight a few points that we hope will be
helpful for your outreach events.


The introductory talk \cite{mason-intro-talk} was extremely challenging, but it interested
the students immensely and its introduction to network theory, its applications, and some of its jargon helped to provide a solid foundation for the interactive modules to follow.  Put another way, the initial challenge helped make life easier for the rest of the event.  We worked with
students from very different backgrounds, but (for the most part) they
were genuinely interested in working with us.  We learned from
our first outreach event that many students learned the key material
in our lesson plans \emph{extremely} quickly (despite the fact that we
discussed topics on which there is active research).  To address this,
we sometimes needed to come up with impromptu activities to fill the
allotted time.  In Module \ref{lesson2} (Small Worlds), for example,
we sometimes discussed material from a different module (such as
Networks and Disease) that we were not running that day. 
When we were able to work with the same students on more than one
module, students usually picked up the key ideas faster in the second module.

Our most successful modules were the most interactive ones, and we
tried to increase the interactivity in the modules as we refined them.
The sessions instigated many interesting and thought-provoking
conversations, and sometimes the best thing we could do was keep our
mouths shut for a while.  In Module \ref{lesson1}, for example,
students used a network-science perspective to think about the importance
of grass at the bottom for survival in an entire food web.  It was
inspiring to see individual thinking, group discussions, and ``ah-ha"
moments without our intervention.
 
Despite the overall success of our efforts, we faced many challenges,
and we haven't yet figured out how to overcome all of them.  We have
gained much more of an appreciation for the difficulties faced by
primary and secondary school teachers than we had before \footnote{Occasional
outreach activities are wonderful, but we wouldn't want to do this
every day!}. For example, sometimes it was difficult to get students to
choose a module in the first place. (We preferred, when possible, not
to choose modules for them.)  Sometimes the majority of students
picked the same topic, and we had to balance group sizes 
with the students' interests.  Many
students were hesitant to participate, and we sometimes had to
tailor the module format to encourage more group-based answers rather
than individuals ones.  Some students were 
disruptive, though teachers from most of the schools were very helpful for managing these situations and otherwise maintaining order.  (This became a serious problem at only one school.)  It is worth remarking that we often noticed a difference in student comfort level 
when they visited us versus when we visited them.
It is also important to note that the smoothest outreach days are not necessarily the most important ones to undertake.  For example, our outreach activities were more difficult when we worked with weaker students, but this type of struggle is a \emph{necessary} one.  We welcome ideas for how to
improve our outreach activities to make them more accessible for a wide variety of students and schools.


\section{Why Bother?} \label{sec6} 


Network science offers an exciting supplement to standard mathematics
curricula in schools.  Spending time on it does take time away from
the existing school curriculum, but we think that it is time well
spent. Examining problems in networks demonstrates that mathematics is
about ideas and abstractions rather than just calculations, and it does
so in a way that is highly visual and closely connected to students'
everyday experiences.  Working through problems like those in our
modules helps students (and teachers) to see a side of mathematics
that is different from most of what they have experienced.  We
emphasize intuition, modeling, and problem-solving, and this provides
a nice complement to the 
calculations to which students are accustomed.  Because of the ubiquity of networks in
everyday life, network science is an ideal subject for these kinds of outreach efforts.  Once we explain to students what a network is, they---just
like professional scientists---see them everywhere they look, and
hopefully they will also start to see mathematics everywhere.

Based on our discussions with the students (and their teachers) and the comments in the survey
forms they filled out for the events in Oxford, the students with whom
we have worked certainly seem to have gained a significant interest in
networks and mathematics.  They viewed the problems that we presented as
puzzles to ponder rather than rote calculations to get out of the way.
It is also a positive experience for students and teachers to have
direct contact with professional mathematicians and scientists to get
a better idea of what it is that we do (which is to try to \emph{solve
  problems}, just like the students were doing).  It is also good to
promote a view of scientists as regular human beings, which
contrasts sharply with what often seems to be the case.
 Meanwhile, the outreach activities help improve the communication skills
of the instructors, and that too is a significant boon.

We target our outreach activities for students of age 13--16, as the younger students in this range have still not decided what subjects to study at university.  (Students specialize very early in the United Kingdom.) We hope to persuade more students to pursue mathematics and science, and we hope that those who don't pursue those subjects will at a minimum gain a greater appreciation of mathematics as well as its importance.  Network science is an eminently accessible subject---that is why many professionals enjoy it, after all---which makes it perfect for these kinds of outreach efforts.


\section{Conclusions and Outlook} \label{sec7}

We are continuing to conduct outreach activities across England, but the only way to make a really big impact is if these activities spread far and wide.  (Let's turn this into a social contagion, okay?)  We hope that we have wet your appetite to conduct outreach activities in schools in your area, and we encourage you to use, steal, adapt, and improve any material in this article or in the SOM.

We recognize that many of you will have different opinions as to what should constitute the contents of modules, and we have already seen that the outreach activities and modules need to be adaptable from one school to another. Please send us your new modules and your improvements to our modules so that we can develop a large repository for networks-related school activities. We are also happy to provide advice or send you less formal notes if you think that they can be helpful.

We also encourage school teachers to include network science in their lessons and math/science-club projects (and, of course, to use our materials and to contact us for any desired discussions or feedback).  We have written only a few modules, but we hope that they convey the hugely important role that networks have the potential to play in getting school students interested in mathematics and science. This is a big challenge, but it is also an exciting opportunity.  

We hope that we have convinced you to engage in outreach efforts in your neighborhoods.  It is a valuable use of your time, and it can have a very large impact. Please don't hesitate to contact us if you would like to discuss this further.


\section*{Acknowledgements}

MAP acknowledges University of Oxford for twice funding these outreach projects as a component of their Pathways to Impact grant from the EPSRC. MAP was also supported by the FET-Proactive project PLEXMATH (FP7-ICT-2011-8; grant \#317614) funded by the European Commission as well as EPSRC grant EP/J001759/1.  MAP and MPR acknowledge a research award (\#220020177) from the James S. McDonnell Foundation. MBD acknowledges support from the EPSRC under the Mathematics underpinning the Digital Economy programme and from the James S. McDonnell Foundation 21st Century Science Initiative - Postdoctoral Program in Complexity Science/Complex Systems-Fellowship Award (\#220020349-CS/PD Fellow). HAH acknowledges support from Michael Stumpf's Leverhulme Trust Grant (\#F/07 058/BP).  LMK acknowledges support from the Natural Sciences and Engineering Research Council of Canada (CGSM \#403601-2011) and TERA Environmental Consultants. We thank Daniel Kim for assistance in developing some materials and Amy Crosweller, Karen Daniels, Lucas Jeub, Sang Hoon Lee, Jes\'us San Martin, and Stan Wasserman for helpful comments on this manuscript.

We thank Somerville College for supplying room facilities free of charge on several occasions and Amy Crosweller of Somerville College for helping to arrange the events.  We our particularly indebted to the numerous volunteers who helped design and/or run modules.  These include Ross Atkins, Senja Barthel, Andrew Elliott, Pau Erola, Virginia Faircloud, Martin Gould, Lucas Jeub, Sang Hoon Lee, Peter Neumann, Sofia Piltz, Jes\'us San Mart\'in, Andrew McDowell, Atieh Mirshahvalad, Ioannis Psorakis, Marta Sartzynska, Karin Valencia, and Kerstin Weller.  (We apologize to anybody who we forgot to mention.)  During events at Somerville College, we also received help from Zoe Fannon, Martin Griffiths, Stanislav Kasjalov, Jess King, Jen Kitson, Josie Messa, Steve Strogatz, and Almat Zhantikin.  Wiesner Vos and Philip Clarkson of Google helped during a panel discussion about mathematics careers.  MAP thanks Chris Budd, Amy Crosweller, Martin Griffiths, and Si\^{a}n Owen for helpful advice in designing this outreach program.


\appendix

\section{Module 1: Networks and the Environment}\label{lesson1}

In this module, which was designed by Laura Keating and Puck Rombach, the students examine how climate change can affect animal species. The students construct and analyze food webs, and they investigate how the extinction of one species can affect other species. 

We start the session by informing the students that they have been hired by the government as applied mathematicians to advise how climate change might affect species survival in the Arctic. Many of the students already possess some knowledge of food chains from their biology lessons, and we build on this to present food webs as networks rather than chains. To help introduce background material, the students first fill out a worksheet that includes questions about what a food web is and how one might construct a food web as a network.  These initial questions also introduce relevant network properties, such as directedness of edges to describe unidirectional energy flow between species.

To further grasp the concept of a food web as a network, we construct a network using a small set of familiar species (e.g., grass, a mouse, an insect, a small bird, and a large bird). We then explore what happens if one or more of those species becomes endangered.  We begin to discuss the potential for extinction cascades and the implications that network structure (e.g., high in-degree or out-degree) can have for an entire ecosystem. 

The final part of the module is spent on a `game' that explores food webs in Arctic ecosystems. We split the students into two groups and give each group a set of Arctic species that are either aquatic or terrestrial.  Both sets of animals include the polar bear, which connects the aquatic and terrestrial food webs.  See the Supplementary Online Material (SOM) for further details and the handouts that we give to the students. Each group of students constructs a network from their respective species, and they write down the adjacency matrix representation of it.  We then ask the students what happens if one of the nodes is removed (i.e., if a species becomes extinct).  Even these small networks are able to help illustrate the complicated and far-reaching effects of species extinction.  At the end of the module, the students present their findings---including which species are most vulnerable, how climate change can affect food webs, and potential ways to reduce the negative effects of extinction.

 
\section{Module 2: Social Networks and Small Worlds}
\label{lesson2}

This module, which was designed by Mariano Beguerisse D\'iaz and Pau
Erola, illustrates that social networks are everywhere and introduces
some of their features.  We examine the notion of a small world---the
introductory talk included the jargon and a picture of a
Watts-Strogatz network---using several examples that we hope are
poignant for the students.

At the beginning, we ask the students whether they have heard of the
idea of ``six degrees of separation''.  The term itself tends to be unfamiliar to most students, but many of them recognize the idea as a familiar one once we explain the jargon.  We discuss Stanley Milgram's package-passing experiments and how it reached a more mainstream
audience through venues such as \emph{The Oracle of Kevin Bacon}
\cite{bacon}.  We play the Kevin Bacon game with the students to try
to find paths between movie actors so that they can find their own
``surprisingly" short paths (e.g., a short path between an actor in a
horror film and one in a children's movie), and we also think about
trying to navigate networks. See the SOM for an example of a handout used for this activity.

We ask the students to explain what is meant by a ``social network"
and to give examples (and to indicate the nodes and edges).  We expand
on the Kevin Bacon game by examining short paths and navigation in
social networks.  For example, we might ask the students ``How many
degrees of separation are there between you and the Queen of England?"
We also tried other famous people, and students on two occasions had very short paths to Nelson Mandela (length 1 in one case and length 2 in another).  We again find short paths
and try to reconcile these close connections with the fact that famous people
are supposedly distant and unreachable.

We discuss who in a social network might help lead to the presence of
many short paths, introduce the notion of hubs, and ask the students to consider this in the context of their Facebook friendships \cite{Ugander2011, Backstrom2011}.
We stress that the notion of a small world applies throughout a network and that many short paths go through hubs.
We also discuss different ``communities" of people and which types of
connections might serve as ``shortcuts" between different communities
rather than ``local" connections within a community.  Typical examples
of the former arose via friendships from vacations or a summer spent
abroad, and naturally the latter tended to be people---classmates,
neighbors, etc.---in close geographical proximity.  

As we developed this module, we added more exercises for the students,
as early versions of the module were sometimes less interactive than
we would have preferred.  One exercise, which did not include a
worksheet, concerns how Facebook chooses which friendships it
recommends to its users.  We encouraged the students to develop their
own recommendation algorithms, and we discussed their ideas on a
whiteboard or flip chart.  In another (very successful) exercise, we
returned to movies and considered the social networks of characters in
movies such as \emph{Toy Story} (see the SOM and \cite{galaxy}). We gave the students a handout with an unlabeled version of the \emph{Toy Story} network, and we asked them
to identify which nodes represent the protagonists and other
characters.  We also asked the students to identify some communities of toys
that had things in common.


\section{Module 3: Networks and Disease}
\label{lesson3} 

This module, which was designed by Heather Harrington and Mariano
Beguerisse D\'iaz, illustrates how thinking about networks arises
naturally when one tries to understand how diseases spread and how to develop good strategies to contain them.

We start this module by asking students to discuss the main
characteristics of an infectious disease and to think about how it
might spread (e.g., what is the main difference between a
non-infectious disease and an infectious one). We encourage the
students to think about a fictional disease that can only spread by
shaking hands and to discuss
 how it might spread in their school.  This quickly leads to the notion of 
disease spread along social networks in schools, and we sometimes combined this module with Module 2 (on social networks and small worlds).  For this discussion, it can be useful to distinguish
online and offline social networks (and also to distinguish between
the spread of ideas and rumours versus diseases).

We also briefly discuss what other types of networks (e.g.,
transportation networks or trade networks) might be important for
understanding, containing, and preventing diseases. We steer the
discussion towards how different types of network topologies can
affect disease spread and vaccination strategies.  (One can ask
similar questions in Module 2.)  If it is too expensive to vaccinate
everybody, then who should be vaccinated?  Some students brought up
node labels in this discussion---one rebellious student proposed that the youngest people should be vaccinated because they (supposedly) had the longest left to live---though our primary focus
was on network topology and how it affects disease spread. In some sessions, students brought up
air travel, which led to a discussion of how such travel has changed
the ways diseases spread.  Students also pointed out that some diseases
could be spread by insects like mosquitos, and we used this
opportunity to introduce bipartite networks and to explain how
vaccination strategies differ in this situation. For example,
fumigation can eliminate many mosquitoes (reducing the number of one
type of node) and slow down the spread of a disease.

\begin{figure}[!t]
  \begin{center}
    \includegraphics[width=0.7\textwidth]{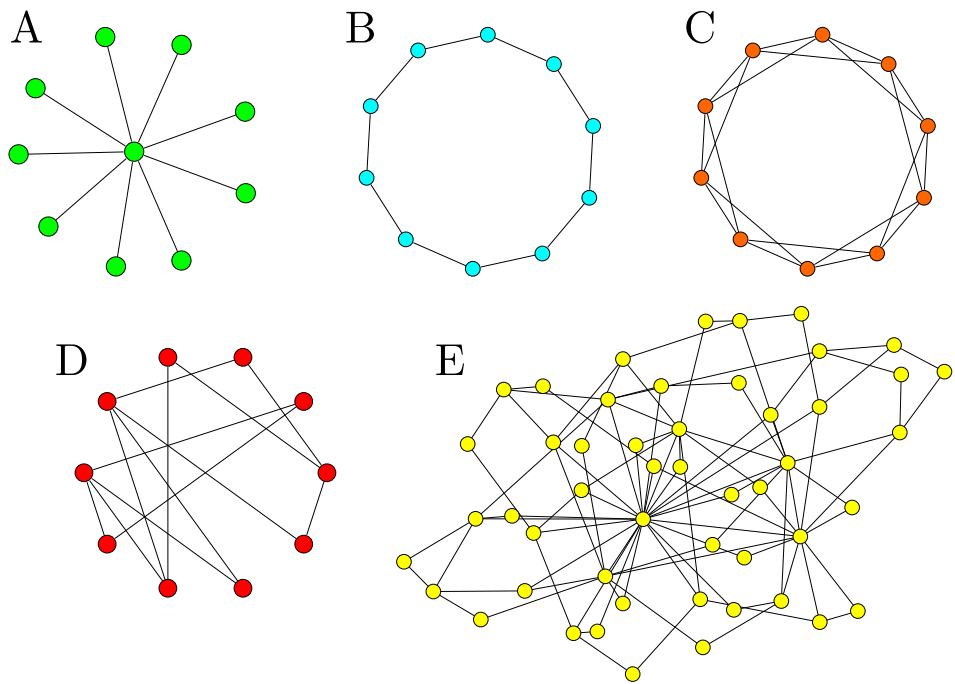}
  \end{center}
  \caption{Example networks for which one can apply different vaccination strategies (see the SOM). ({\bf A}) star network, ({\bf B}) circular lattice, ({\bf C}) regular circular
    lattice with four neighbors ({\bf D}) Erd\H{o}s-R\'enyi random
    graph, and ({\bf E}): Barab\'asi-Albert network.}
  \label{fig:vaccines}
\end{figure}

The largest portion of the module is a hands-on activity in which we
distribute handouts (see the SOM) with various example networks (see Fig.~\ref{fig:vaccines})
to the students and ask them to devise possible vaccination strategies in each case if they are only
allowed to vaccinate three or fewer nodes.  The students realized
quickly that this question was much harder to answer for some network
topologies than for others.
This was an interesting point of discussion, as it
allows the students to consider how one might develop a vaccination strategy in
real networks, which are much more complicated.  
Moreover, given that one needs to think about the
answer even if one knows network structure exactly, we can discuss how to develop strategies when some (or even a lot) of the network structure is not known.
We ask the students what would they do if we only know
a network has a particular structure (e.g., suppose that one knows that it was generated using a Barab\'asi-Albert mechanism)
but do not know anything else. This question generated a lively discussion. A useful hint for many students is to ask what would happen if we choose a node at random and then ask him/her to
choose a friend to vaccinate (rather than vaccinating the original
node).  We also sometimes discuss the time-ordering of contacts in
social networks and how that can influence disease spread.

In addition to discussing diseases specifically, it can be useful to
encourage the students to think about other contexts in which ``vaccination" strategies might be useful.  One key question is the difference between the spread of an idea and a disease, and one
might also wish to discuss other dynamical processes on networks.

This module is particularly nice for illustrating that mathematics
shows up in many situations that the students (and their teachers) did not previously consider to be mathematical.  This occasionally came up in discussions of viable careers for people who study mathematics at the university
level, and we highlighted that nowadays mathematicians work closely
alongside health professionals.


\section{Module 4: How Google Works}
\label{lesson4}

This module, which was designed by Mariano Beguerisse-D\'iaz, Sang
Hoon Lee, and Lucas Leub, aims to introduce Google's PageRank
algorithm for ranking pages on the World Wide Web
\cite{brin98,newman2010}.  (Daniel Kim also assisted in developing materials for this module.)

To start the module, we ask the students to imagine a world without Google or other search engines and to develop their own strategies for finding information on the Web.  Usually, one of the first ideas is to compile an exhaustive list of every Web page.  We use this to introduce the idea of a crawler 
to navigate Web pages, and this leads naturally to the notion of representing the Web as a network with directed edges (the hyperlinks) between nodes (the Web pages).  

We ask the students to think about how to figure out whether a Web page is relevant for the information one seeks, and this leads almost immediately to the issue of how one should rank Web pages in order of importance. One possibility that the students
quickly bring up is that one can develop rankings based on the textual content
of a page.  (In one case, we had a good discussion about how we would
try to use an automated method to distinguish the Amazon rain forest
from Amazon.com.) We let the students know that the first Web search
engines used to be ``curated'' by hand, which limited how much of the
Web could be explored. This limitation was an incentive to people to
seek algorithmic methods to rank pages, which is what we want the students to explore.  We ask the students to develop ideas for how to use the network structure of the Web to rank
pages.  (The difficulty of this transition in the discussion varied strongly from one to school to another.)
Most of the time, the first structure-based ranking that the students propose is to
rank Web pages according to the number of incoming hyperlinks (i.e.,
according to in-degree).  We discuss whether someone can cheat this
system (as well as simple text-based systems) to improve the ranking of a page and whether better methods are available.  

\begin{figure}[!t]
  \begin{center}
    \includegraphics[width=0.4\textwidth]{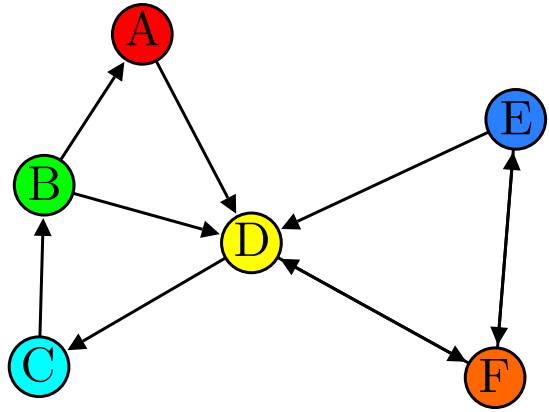}
  \end{center}
  \caption{An example of a directed network whose nodes we ask
    students to rank in order of importance. This network is strongly
    connected (so any Markov chain on it is ergodic), so we can ignore
    the problem of dead-end nodes (i.e., nodes with an out-degree of
    $0$).  However, we did discuss the notion of dead ends on many occasions
    that we ran this module. The ranking of the nodes in this graph
    from largest to smallest PageRank score (in parenthesis) is as follows: D (0.3077), F (0.2051), B (0.1538), C (0.1538), E
    (0.1026), and A (0.0769).}
  \label{fig:google-graph}
\end{figure}

In Fig.~\ref{fig:google-graph}, we show an example network that we
use to help guide our discussion of how to rank Web pages.  This
example is particularly useful for moving beyond ideas for ranking based on
the text on a page, as we can pretend that no such text
exists (or that it is otherwise impossible to distinguish Web pages
based on their text).  We give the students a
handout with this network (see the SOM), and we
ask them to rank the nodes in order of importance (and also to
indicate how they have defined ``important").

Motivated by the fact that people often seem to explore the Web by
``randomly" following hyperlinks---who hasn't done this on Wikipedia?---we ask the students whether they can develop a ranking method based on
this idea. 
We use phrasing along the following lines: If we
have a large number of monkeys---it is very compelling to refer to
random walkers as ``monkeys" \cite{bcs-press}---who are clicking on
hyperlinks randomly, what ranking would we obtain based on the number
of times each page is visited.  It can also be useful to discuss why Wikipedia is a ``monkey trap", in the sense that many Wikipedia pages have high rankings in Google searches. (We occasionally discussed having one random walker versus having a large number of random walkers.)  Using
these questions, we introduce the rationale behind PageRank.
\emph{Crucially}, we try to avoid words like ``eigenvalue" and
``eigenvector" (and ``ergodic", ``Markov", etc.), though we do attempt to get the students to compute (by hand) the PageRank eigenvector for a network like the one
in Fig.~\ref{fig:google-graph}.  They just don't know that what they are computing is called an eigenvector.

The example network in Fig.~\ref{fig:google-graph} is very
instructive.  Different choices for how to measure node importance
(e.g., in-degree versus out-degree) lead to different rankings, and we
have interesting discussions regarding which ranking is ``correct".
(These ideas could also be used to develop a module that focuses on
centralities more generally---e.g., intuition related to the notion of
``betweenness" sometimes comes up in Modules 2 and 3---as well
as how one might change the notion of importance depending on the
question one wants to answer.) An interesting feature of the network
in Fig.~\ref{fig:google-graph}, which is worth asking the students to
try to prove, is that nodes B and C have the same PageRank score.

To compute the rankings, the students count the number of times the nodes are visited on different walks through the network.  We and the students use two primary techniques for this calculation: (1) start from a uniform distribution and iteratively count the number of walkers on each node, or (2) try to identify relative orderings for the ranking of different nodes without calculating individual probabilities. The maximum out-degree in the example network is 2, which makes it easy to simulate a random walk by flipping a coin to decide which edge to follow.  The students soon realize that one can get to node D from almost
everywhere, and that it is indeed the most visited node in a (conventional) random walk.  This then makes it the most important node in this ranking scheme. The students then realize
that the nodes receiving edges from it (C and F) must come next in the rankings, and they discuss how to break the tie. Students also note that C and B are always visited the exact same number of times (as long as we don't stop the walk before a monkey has had the chance to leave C).  
In some cases, we were able to get the students to calculate the actual percentage of visits for each node rather than only determining the rank order.

When there is enough time, we discuss that a monkey gets ``trapped'' on a Web page with an
out-degree of $0$. This problem can be illustrated by adding a
dead-end node G to the network in Fig.~\ref{fig:google-graph}. We
ask the students to think about how they can change their ranking
methodology to be able to deal with this situation.  This gives the opportunity to discuss
the idea of a random walk with ``teleportation" (e.g., at each node,
one follows an edge with a probability $p$ or otherwise chooses some
other node in the network via a random process).  Once the dead-end node has been added, the graph is no longer ``strongly connected", which we can use to illustrate that even a small perturbation of a network can change its properties in a fundamental manner.

As part of this module, we sometimes discuss clever scientific ways to use Google---such as trying to measure the similarity between two football players by examining how often they show up together in Google searches \cite{googlingsocial}.


 
 \section{Module 5: Introduction to Graph Theory}\label{lesson5}

This module, which was designed by Puck Rombach, provides an introduction to graph theory through the discussion of some famous mathematical ``puzzles".  In contrast to the other modules, it focuses on the theoretical (or ``pure") side of mathematics.  It is also arguably our most popular and successful module.

Graph theory's deep connection with networks makes it an area of pure mathematics that can resonate with students. The problems that we discussed with the students happen to also be relevant for applications (and we mention them when we are asked about them), but that is not our focus.  We want to show the students how to formulate and prove theorems, and we want to convey our excitement for and the value of abstract mathematical ideas for their own sake.  This module also illustrates that university mathematics need not require any numbers or calculations, which is an important difference from the kind of mathematics with which students are familiar from schools.

\begin{figure}[htp]
  \centering 
  \includegraphics[width=60mm]{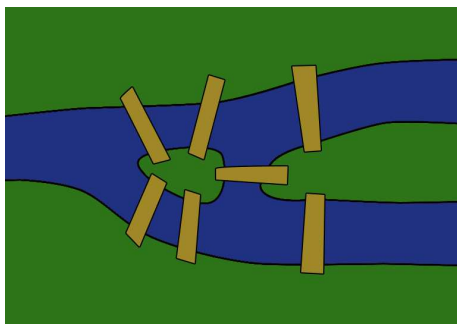}
  \caption{\emph{The Bridges of K\"onigsberg.} Is it possible to create a walk that crosses each bridge exactly once and also returns to its starting point?
  }\label{fig:kon}
  \end{figure}

\begin{figure}[htp]
  \centering 
  \includegraphics[width=60mm]{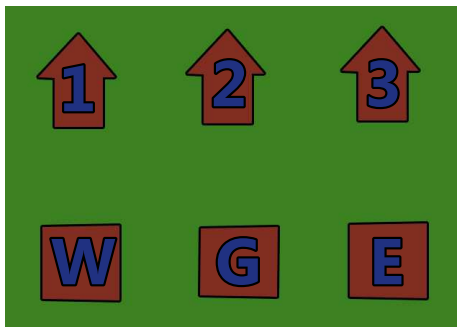}
  \caption{\emph{The Water, Gas, and Electricity Puzzle.} Is it possible to connect every house to all of water, gas, and electricity such that no lines cross?
  }\label{fig:wge}
  \end{figure}

Graph theory is wonderful because its basic ideas and many of its interesting problems can be explained in a few minutes in a way that allows people with little or no mathematical background (such as young children) to understand them.  Two examples that we discuss with the students are the Bridges of K\"onigsberg (see Fig.~\ref{fig:kon}) and the Water, Gas, and Electricity puzzle (see Fig.~\ref{fig:wge}).  Both of these puzzles are unsolvable, which one can demonstrate by systematically exhausting all possible solutions.  We ask the students to try to find solutions to these puzzles and then, when they can't, to try to prove that no solutions exist.

Excitingly, both puzzles hint at deep and general theorems. The Bridges of K\"onigsberg~\cite{euler} is unsolvable because it is both necessary and sufficient that every node have an even degree for a graph to have an Euler cycle (i.e., a circuit that runs along every edge exactly once and returns to its starting point).  In this case, no nodes have an even degree. Proving the necessity is easy---every time that an island (i.e., node) is ``entered", it must also be exited---but proving sufficiency is more difficult.  This problem also allows us to discuss the notion of necessary and sufficient conditions for results to be true.
The Water, Gas, and Electricity puzzle is unsolvable because $K_{3,3}$ (the complete bipartite graph on two sets of three nodes) is not planar \footnote{A ``planar" graph is a graph that can be drawn on a plane such that no edges cross each other.}.

 Another family of problems that we discuss at length with the students concerns graph coloring.  Graph-coloring problems are easy to explain, but they can be excruciatingly difficult to solve.  The most famous graph-coloring result is the Four-Color Theorem \cite{diestel}, which we explore in detail with the students.  We ask them the following question: How many colors do we need to color a map such that two countries that share a border are not colored with the same color?  (Maps can be represented as planar graphs and vice versa, so map coloring is the same as planar graph coloring.) We provide paper and markers and examine the colorings that the students draw.  This allows us to determine quickly if they understand the definition of ``coloring".  We keep challenging them to try to construct maps that require more colors (and we typically need to make an explicit statement that we are disallowing islands).  The students find through empirical observation and discussion that none of them ever need more than four colors, and we eventually let them know that it impossible to construct a planar map of contiguous countries that isn't four-colorable. \footnote{This was proved in the 1970s by Kenneth Appel and Wolfgang Haken \cite{app70s}, but it required exhaustive computer searches and hundreds of pages of analysis. There isn't (yet) an elegant proof for of this theorem.}  In one memorable incident, a student insisted (despite our statement that it was impossible) that he was going to construct a map that couldn't be four-colored, he kept working on that for the rest of the day, and then he insisted as the students were leaving that he was going to continue working at it and get it to work.  He'll probably be a famous mathematician someday, as he's already got the right persistent attitude for it.

\begin{figure}[htp]
  \centering 
  \includegraphics[width=60mm]{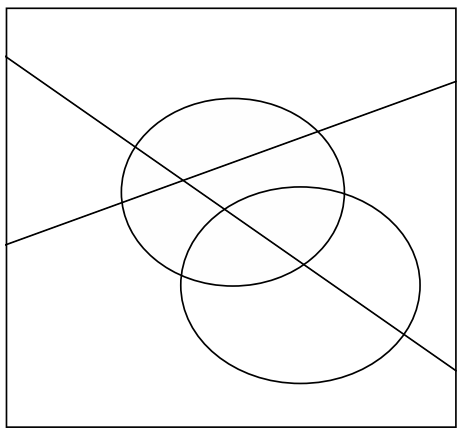}
 \caption{An infinite-border map, for which one can prove a Two-Color Theorem.} \label{fig:2col}
  \end{figure}

To get the students to get their hands dirty with mathematical proofs, we then discuss a much simpler result that we call a \emph{Two-Color Theorem}.  We consider maps with the special rule that borders cannot end.  That is, every border must either go off of the page in 
all directions (imagining that it goes to infinity) or it has to connect to itself in a cycle.  To make things simpler (although the result holds without this simplification), we do not allow borders to cross themselves. In Fig.~\ref{fig:2col}, we show an example of this special ``infinite-border" map (IB-map). We teach the students how to prove this Two-Color Theorem using induction~\cite{gardner}, which we explain is a common approach in mathematics.  (The proof is in the SOM.)

If there is extra time or if it is useful to discuss something different, we also go through the proof of Hall's Marriage Theorem.  (This problem can also be discussed without bringing up marriage, though that is the traditional setting of the problem.) This theorem gives a necessary and sufficient condition for marrying each member of one group of people to a member of another group of people, given that some of the latter have conditions that limit who they are willing to marry.\footnote{Conveniently, half of the 2012 Nobel Prize in Economics was awarded for theoretical work on generalizing HMT \cite{nobel2012econ}.  We used this as part of our response to a skeptical question about HMT ever being useful---the reaction to our answer was wonderful---though we also stressed that that didn't matter, as we were only concerned with the fact that proving the theorem is interesting.} The setting for HMT is enjoyable to explain, and it usually elicits a few laughs from the students. 

Student reaction to this module has been overwhelmingly positive.  To our great pleasure, most students have enjoyed solving problems for their own sake without having to be encumbered by any external importance. Most of the questions are very challenging and difficult to answer straight away.  When running this module, we encourage the students to think like mathematicians. If they find a problem to be too hard, then we encourage them to consider a simpler but related question that they can try to answer first.  If students doesn't know an answer, it is good to ask them to indicate what they \emph{do} know.  For particularly keen students, of course, this module can be scaled up to make it more challenging.

 
 \section{Module 6: Do Your Friends Have More Friends Than You Do?}\label{lesson6}



Two decades ago, Scott Feld wrote a well-known paper that discusses why your friends have more friends than you do \cite{feld}.  Because this can be explained using simple mathematical arguments on some networks (such as the configuration model \cite{newman2010,bollobas}), we decided that this idea would make a very cool module.

This module, which was designed by Lucas Jeub, starts by asking the students whether or not they have heard about this result.  In the one time we ran the module, none of the students had heard about anything like this (which isn't terribly shocking), and we actually want students to express skepticism about this kind of result. Our plan in such a situation is ask the students to come up with an argument of why such a result can't be true followed by guiding them through the intuition for the correct result that the mean number of friends of a friend is larger than one's mean number of friends.

To illustrate this result, we ask the students to think about their Facebook friends and the number of friends of their most popular Facebook friend.  
We also thought about getting the students to write down the number of people in the room that they consider to be friends, as this can of course help illustrate the same phenomenon.  However, you can already probably see a major flaw in this attempt to illustrate the key idea: the students were rather shy about saying how many friends they had.  

We used the configuration model as a toy situation to illustrate the key phenomenon.  
We asked the students to write down their number of friends on a piece of paper.\footnote{As indicated above, it would have been better to use a less personal mechanism to illustrate this example. For example, one can ask students to choose an arbitrary positive integer bounded above by the number of people in the room and to just pretend that that number is their number of friends.}  We tabulated the numbers and asked the students to try and construct a network with the given degree sequence, and the idea of the configuration model than comes naturally by considering the set of all possible networks that one can construct from this information.

We were then hoping to get into a discussion on an intuitive level of whether the fact that social networks are rather different from the configuration model (e.g., because of triadic closure) changes the result, and we thereby wanted to get into some issues regarding the structure of social networks.  However, although we were able to get the students to understand the arguments with the configuration model, our attempts to scale up from this point were unsuccessful.  An even bigger issue is that the calculations necessary to verify the key result are tedious even for small networks. We ended up finding something else to discuss with the remaining time and decided that this module needed to go back to the drawing board.

A while after we ran this module, Steve Strogatz presented an excellent explanation of Scott Feld's result in a \emph{New York Times} article \cite{strogatzNYT}.  We think that his explanation can serve as a useful springboard for a good module, and we suggest starting from there if you like the idea of a module with this theme.  We believe that this theme has the potential to make an excellent module.

 
 \section{Module 7: Structural Balance (The Ad Hoc Module)}\label{lesson7}

This module wasn't ``designed" by anybody, though we ran it once on an ad hoc basis instead of Module 6 when the latter hadn't worked out as well as we would have liked earlier in the day.

We needed something at the last minute, as trying Module 6 without
going back to the drawing board first was not an option.  Motivated by
a brilliant colloquium by Steve Strogatz the previous day, we decided
that structural balance would make an excellent topic.  We covered the
basic ideas behind structural balance, including whether or not three
mutually antagonistic connections in a triad should be considered as
balanced.  We also guided the students through work by
Cartwright and Harary \cite{harary} and asked them to consider this in
real life as well as on Massively Multiplayer Online Role-Playing
Games (MMORPGs) \cite{szell}.  (For some students, it might not be
entirely clear which of these is more real.)

In our original brainstorming for sessions to design, we had actually
considered developing a module on structural balance.  The ad hoc
module based on that idea was surprisingly successful given the lack of formal design and preparation, so we feel that structural balance would make a good module to develop more carefully.  One can of course bring up alliances and conflicts in war and among
schoolmates, and one can also discuss games like Risk in addition to
MMORPGs.



\end{document}